\documentstyle[twoside,fleqn,espcrc2]{article}
\input epsf
\def\BNL {Department of Physics, Brookhaven National Laboratory, Upton, NY 11973\\}
\newcommand{\LL}{\left\langle}
\newcommand{\RR}{\right\rangle}

\newcommand{\BE}{\begin{equation}}
\newcommand{\EE}{\end{equation}}
\newcommand{\BEA}{\begin{eqnarray}}
\newcommand{\EEA}{\end{eqnarray}}

\newcommand{\etal}{{\em et al.\ }}

\newcommand{\gbeta}{6/g^2}
\newcommand{\CS}{{\rm SU(N_f)_L\times SU(N_f)_R}}

\def\simge{
    \mathrel{\rlap{\raise 0.511ex
        \hbox{$>$}}{\lower 0.511ex \hbox{$\sim$}}}}
\def\simle{
    \mathrel{\rlap{\raise 0.511ex
        \hbox{$<$}}{\lower 0.511ex \hbox{$\sim$}}}}

\newcommand{\AmS}{{\protect\the\textfont2
  A\kern-.1667em\lower.5ex\hbox{M}\kern-.125emS}}

\title{ Lattice QCD with domain wall quarks and applications
to weak matrix elements }
\author{T. Blum\thanks{email: tblum@penguin.phy.bnl.gov} 
and A. Soni\thanks{email: soni@penguin.phy.bnl.gov}
\address{ \BNL } }

\begin{document}

\begin{abstract}
Using domain wall fermions, 
we estimate $B_K(\mu\approx 2\,{\rm GeV})=0.628(47)$ in quenched QCD which 
is consistent with
previous calculations. At $\gbeta=6.0$ and 5.85 we find
the ratio $f_K/m_\rho$ in
agreement with the experimental value, within errors.
These results support expectations that $O(a)$ errors are
exponentially suppressed in low energy ($E\ll a^{-1}$) observables, and 
indicate that domain wall fermions have good scaling behavior at relatively
strong couplings. We also demonstrate that the axial current
numerically satisfies the lattice analog of the usual 
continuum axial Ward identity.
\end{abstract}
\maketitle

\vskip -.2 in

A basic feature of the strong interactions has been 
missing in lattice calculations, the $\CS$ chiral flavor symmetry of the
light quarks. 
We recently reported\cite{US} on calculations
using a new discretization for simulations
of QCD, domain wall fermions
(DWF)~\cite{KAPLAN,SHAMIR}, which preserve chiral symmetry on the lattice
in the limit of an infinite extra 5th dimension.
There it was demonstrated
that DWF exhibit remarkable chiral behavior\cite{US} even at relatively large
lattice spacing and modest extent of the fifth dimension.
Here we give further results using DWF which are of
direct phenomenological interest\cite{US2}.
 
In addition to retaining chiral symmetry, DWF are also
``improved'' in another important way. 
In the limit that the number of sites in the extra dimension, $N_s$, goes 
to infinity, the leading discretization error in the
effective four dimensional action for the light degrees of freedom goes like
$O(a^2)$. 
This theoretical dependence
is deduced from the fact that the only operators available to
cancel $O(a)$ errors in the effective action
are not chirally symmetric. 
For finite $N_s$, $O(a)$ corrections are expected to be exponentially 
suppressed with the size of the extra fifth dimension. 
Our calculations for $B_K$ show 
a weak dependence on $a$ that is easily fit to an $a^2$ ansatz.
 
We use the boundary fermion variant of DWF developed by Shamir. For details,
consult Kaplan\cite{KAPLAN} and Shamir\cite{SHAMIR}. 
See Ref.~\cite{SHAMIRandFURMAN} for a discussion of the $4d$ chiral 
Ward identities (CWI) satisfied by DWF.
Our simulation parameters are summarized in Table~\ref{RUNTABLE}. 

\vskip -.2 in
\begin{table}[hbt]
\caption{Summary of simulation parameters.
$M$ is the five dimensional Dirac
fermion mass, and $m$ is the coupling between layers $s=0$ and 
$N_s-1$.}
\begin{tabular}{|c|c|c|c|c|}
\hline
\hline
$\gbeta$ & size & $M$ & $m$&$\#$ conf\\
\hline
5.85& $16^3\times 32\times 14$& 1.7& 0.075&34\\&&&0.05&24\\
\hline
6.0 & $16^3\times 32\times 10$& 1.7& 0.075&36\\&&&0.05&39\\
					       &&&0.025&17\\ 
\hline
6.3 & $24^3\times 60\times 10$& 1.5& 0.075&11\\&&& 0.05&14\\
\hline
\hline
\end{tabular}
\label{RUNTABLE}
\end{table}
\vskip -.2 in
We begin with the numerical investigation
of the lattice PCAC relation.
The CWI are satisfied exactly on any configuration
since they are derived from the corresponding operator identity. We
checked this explicitly in our simulations.
In the asymptotic large time limit, we find for the usual PCAC
relation
\BEA
2\sinh{\left(\frac{a m_\pi}{2}\right)}
\frac{\LL A_\mu|\pi\RR}{\LL J_5|\pi \RR}&=&
2 m +2\frac{\LL J_{5q}|\pi\RR}{\LL J_5|\pi\RR},
\label{ratio}
\EEA
which goes over to the continuum relation for $a m_\pi \ll 1 $ and 
$N_s\to\infty$ (see Ref.~\cite{SHAMIRandFURMAN} for operator definitions).
The second term on the r.h.s. is anomalous and vanishes as
$N_s\to\infty$. It is a measure of explicit chiral symmetry breaking
induced by the finite 5th dimension.
At $\gbeta=6.0$ and $N_s=10$ we find the l.h.s. of Eq.\ref{ratio} to be
0.1578(2) and 0.1083(3) for $m=0.075$ and 0.05, respectively. The 
anomalous contributions for these two masses are $2\times$(0.00385(5) and
0.00408(12)), which appears to be roughly constant with $m$.
Increasing $N_s$ to 14 at $m=0.05$,
the anomalous contribution falls to
$(2\times)$ 0.00152(8) while the l.h.s. is 0.1026(6), 
which shows that increasing $N_s$ really does take us
towards the chiral limit.

Next we investigate the matrix element of $O_{LL}$. At $\gbeta=5.85$ the two
data points extrapolate linearly to -0.0005(100) at $m=0$. At $\gbeta=6.0$
the three data points extrapolate to -.004(9). 
At $\gbeta=6.3$, the two points extrapolate to 
0.05(3). This slight overshoot is not unexpected since the values of
$m$ used here correspond to rather heavy quarks. In our initial study
we found a similar behavior\cite{US}, and 
as the quark mass was lowered, the required linear behavior set in.
All of the above results are for $N_s=10$ except at $\gbeta=5.85$ where
$N_s=14$ was used for reasons explained below.

\vskip -.25in
\begin{figure}[hbt]
    \hskip -.1in\vbox{ \epsfxsize=3.0in \epsfbox[0 0 4096 4096]{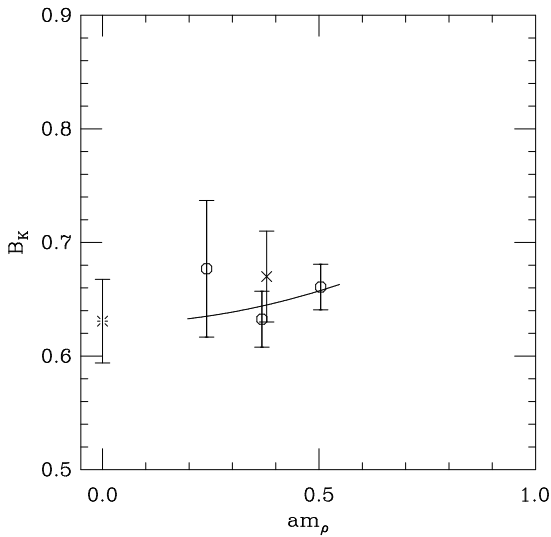} }
    \vskip -.5in
    \caption{The kaon B parameter. 
	The cross (not used in the fit) denotes
	the partially unquenched result using KS sea quarks\protect\cite{COL}.}
    \label{BK}
\end{figure}
\vskip -.25in
In Fig.~\ref{BK} we show the kaon B parameter.
The results for $B_K$ depend weakly on $\gbeta$, and are well fit to a
pure quadratic in $a$.
We find $B_K(\mu=a^{-1})=0.628(47)$
in the continuum limit. 
This value is already consistent
with previous results\cite{BPAR} though
it does not include the
perturbative running of $B_K$ to a common energy scale.
This requires a perturbative calculation to determine the renormalization
of $O_{LL}$, which has not yet been done.
The energy scale at $\gbeta=6.0$ is roughly 2 GeV.
Also, we find $B_K(\mu=a^{-1})=0.67(4)$ on a set of 20
Kogut-Susskind lattices with $m_{KS}=0.01$ and $\gbeta=5.7$\cite{COL} and the
same five dimensional lattice volume as the point at $\gbeta=6.0$. 

At $\gbeta=6.0$,
we have also calculated $B_K$ using the partially conserved axial
current $A^a_\mu(x)$ (and the analogous vector current).
This point split conserved current requires explicit factors of the gauge links
to be gauge invariant. Alternatively a gauge non-invariant operator
may be defined by omitting the links; the two definitions
become equivalent in the continuum limit.
Results for the gauge non-invariant operators agree within small 
statistical errors with those obtained with
naive currents, Fig.~\ref{BK}(see Ref.\cite{SHAMIRandFURMAN,US}
for operator definitions). 
The results for the gauge invariant operators
are somewhat larger: $B_K^{inv}(\mu=a^{-1})=0.857(20)$ and 0.946(28)
at $m=0.05$ and 0.075, respectively. A similar situation holds in the
Kogut-Susskind case where it was shown that the gauge invariant operators
receive appreciable perturbative corrections which bring the two results
into agreement~\cite{ISH}.

Assuming PCAC, the pseudoscalar decay constant is 
determined from the measurement of $\LL 0| J^a_5|P\RR$. 
We find $f_K=159(14)$ MeV 
and 164(12) MeV for $\gbeta=6.0$ and 5.85, respectively. 
The errors are statistical and do not include 
the error in the lattice spacing determination from $am_\rho$.
The central values agree with experiment, $f_{K^+}= 160$ MeV.
The lattice spacing determinations from $am_\rho$ 
are $a^{-1}=1.53(27)$, 2.09(21),
and 3.20(81) GeV at $\gbeta=5.85$, 6.0, and 6.3, respectively. 
Alternatively, we may form the dimensionless ratio $f_K/m_\rho$. We find
for $\gbeta=5.85$ and 6.0, $f_K/m_\rho=0.213(42)$ and 0.206(27), 
where we have added the statistical errors naively in quadrature. The
experimental result is 0.208.
The decay constant calculated directly
from the matrix element of the partially conserved axial current 
agrees with results using the matrix element of the pseudoscalar density,
up to the anomalous contribution which is slightly less than the 
statistical error.
While the above indicate good scaling behavior, they
must be checked further with improved statistics and a fully
covariant fitting procedure. 
More importantly, the continuum limit still has to be taken:
a recent precise calculation using quenched Wilson
quarks by the CP-PACS collaboration gives a value for $f_K/m_\rho$ in
the continuum limit that is inconsistent with experiment~\cite{CPPACS},
so the above agreement with experiment may be fortuitous. 

\vskip -.2in
\begin{figure}[hbt]
    \hskip -.1in\vbox{ \epsfxsize=3.0in \epsfbox[0 0 4096 4096]{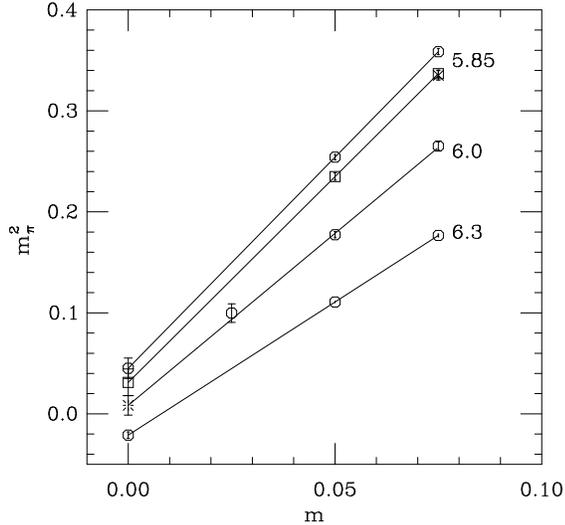} }
    \vskip -.5in
    \caption{The pion mass squared.}
    \label{mpi2}
\end{figure}
\vskip -.25in
In Fig.~\ref{mpi2} we show the pion mass squared as a function of $m$. 
The data at $\gbeta=6.0$ and 6.3 are consistent with chiral perturbation
theory.
At $\gbeta=5.85$ the two masses extrapolate to 0.045(10) for $N_s=10$
and 0.031(13) for $N_s=14$.
This discrepancy is probably not due to higher order
terms in the chiral expansion since the physical quark masses are 
light compared to the masses at the other couplings,
and the curvature would have the wrong sign.
We see a large downward shift in $m_\pi^2$ as $N_s$ goes from 10 to 14. However,
increasing $N_s$ to 18 at $m=0.075$ has a negligible effect(the point at
$m=0.05$ should also be checked since the slope may increase).
We note that the anomalous contribution to 
Eq.~\ref{ratio} is more than double the value at $\gbeta=6.0$.
In the case of the vector Schwinger model, it was
found that topology changing gauge configurations can induce 
significant explicit chiral symmetry breaking effects\cite{VRANAS}. 
Further investigation is required.

Our study shows that DWF are an attractive alternative 
for lattice QCD calculations where chiral symmetry is crucial.
DWF yield good agreement with 
expectations from chiral perturbation theory without the 
complicated mixing of operators required with
Wilson quarks, or the entanglement of flavor and space-time 
degrees of freedom as with Kogut-Susskind quarks. 
Up to exponentially small corrections,
DWF maintain the full chiral symmetry of QCD 
at relatively strong couplings, and thus
should have more continuum-like behavior.
The data presented here seem to indicate just that, though future studies
with improved statistics are needed to confirm this. 
This improved scaling may
compensate for the added cost of the extra dimension.

We thank S. Aoki and Y. Shamir for useful discussions.
This research was supported by US DOE grant
DE-AC0276CH0016. The numerical computations were carried out on the 
NERSC T3E.

%
%
\end{document}